\documentclass[11pt]{article}
\usepackage{geometry}                
\usepackage{graphicx}
\usepackage{amssymb}
\usepackage{epstopdf}
\usepackage{amsmath, amssymb, graphics}
\newcommand{\mathsym}[1]{{}}

\title{Diffraction at the LHC:
a non-technical introduction}
\author{Sebastian White\\
Physics Department, Brookhaven National Lab, Upton, NY 11973}
\date{}                                           

\begin{document}
\maketitle

\begin{abstract}
	In diffractive interactions of protons or nuclei a violent collision can occur that leaves the forward going particle completely intact -with probability determined by the structure of the proton or nucleus.

	At very high energies these collisions also occur with both incident particles remaining intact. This is called central exclusive production. If a new particle, such as the Higgs boson, were produced exclusively this process would give a precise measurement of its mass and test for expected properties of the Higgs. Because of its unusual features this process is also a promising discovery tool.

	In this paper I focus on analogous electromagnetic processes because many aspects apply to both- particularly the role of coherence. Also, topics in
	diffraction with nuclear beams are based on electromagnetic interactions. I also discuss two proposed measurements in ATLAS with Pb beams and with
	proton beams (diffractive Higgs production).

\end{abstract}

\section{2009}

	This year marks the turn on of CERN's Large Hadron Collider (LHC), probably the most complex scientific project ever.
On December 8 CERN transferred $\sim$25 GigaVolt (GeV) protons from the main campus in Switzerland to the SPS in France, where they were accelerated
to 450 GeV and used to fill the 2 colliding beams of the LHC. As they circulated in the 27 km circumference LHC tunnel the protons were further accelerated and reached a
2,360 GeV collision energy in ATLAS- a new energy record.

\begin{figure}
\centering
\includegraphics[width=0.5\linewidth]{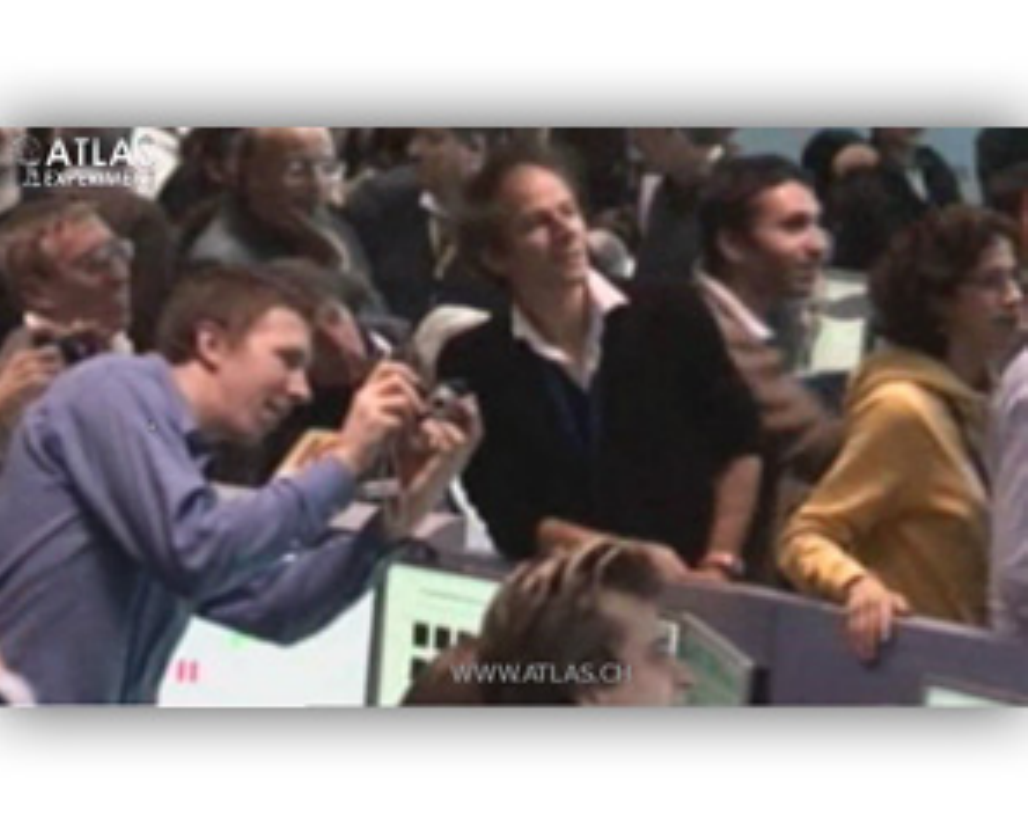}
\caption{First arrival of beam after 20 years of preparation in ATLAS.}
\label{fig:beam}
\end{figure}

	This year also marks the 100th anniversary of the atomic model- based on results from scattering of
$\alpha$-particles from metal foils. In the 100 years since Rutherford's discovery we have focused in on the atom by a factor corresponding to reducing
the distance from New York city to Mazatlan to the size of a thumbnail (ie. our current limit on the radius of the electron or the quarks). The LHC will certainly
continue this trend.

	The LHC was designed to improve understanding of particle interactions and the structure of matter. It 
addresses questions about our picture of elementary particles, including the Higgs. For example, it now seems likely that other types of particles, far more abundant, will
be found as suggested by the dark matter puzzle. It also addresses questions about space and time-- such as whether there are 
additional spatial dimensions beyond our current resolution and whether space is continuous or discrete.
	
\begin{figure}
\centering
\includegraphics[width=0.4\linewidth]{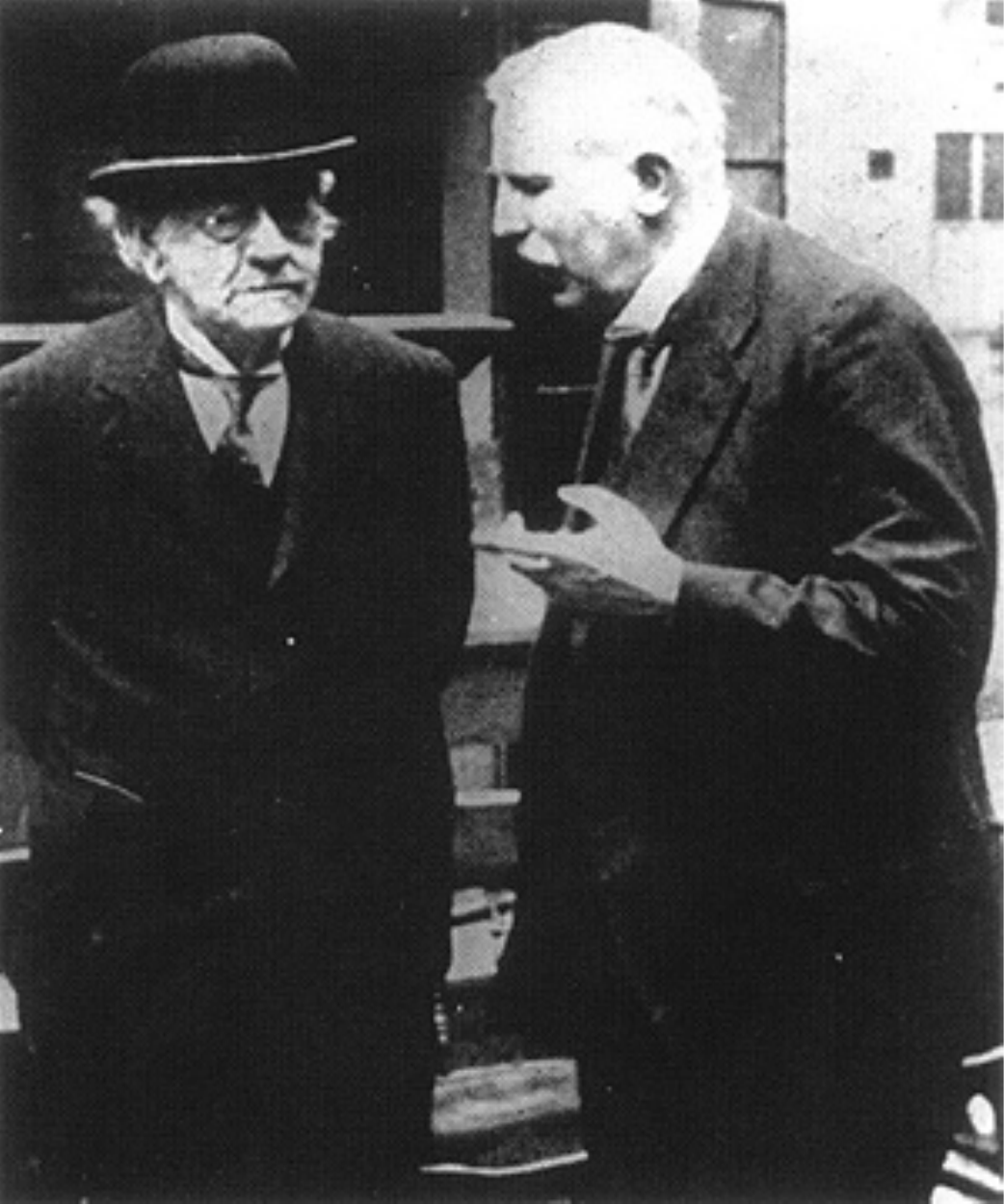}
\caption{Rutherford's teacher, J.J. Thomson had discovered the electron 12 years earlier. }
\label{fig:thompson}
\end{figure}

\section{Elastic Scattering}
	
	The measurements by Geiger and Marsden in Rutherford's lab at Manchester used a collimated beam of $\alpha$-particles from Radium-C decay (
kinetic energy $\sim$ 5 MegaVolt(MeV) ). The beam passed through a gold foil and scattered $\alpha$'s struck a screen coated with a thin layer of Zinc Sulfide(ZnS) powder. ZnS
scintillates when energy is deposited in it and the light from one $\alpha$ can be seen by eye. 

	In the, then prevailing, model of the atom, the plum pudding-due to Thomson, it would be very difficult to scatter through an angle of -say- 10 degrees. Instead a considerable rate was found at large angles. In Rutherford's paper he shows that this rate couldn't be accounted for by random walk through many small scatters. Instead the angular dependence $\sim\sin^{-4}{\theta/2}$ was exactly what you would expect if all of the mass and charge of the nucleus was concentrated at one point.
	
	Rutherford calculated the distance of closest approach of $\alpha$'s based on the repulsive potential of the gold nucleus and found a value of  $\sim 15-20\times10^{-13}$cm-
tantalizingly close to what we now know is the gold nucleus size. However, measurements continued after the war and other, lighter, elements (at Moseley's suggestion) were also tried. In a 1927 paper with Chadwick\cite{Rutherford} on $\alpha$ scattering from Helium, Rutherford found evidence for a ``region of abnormal interactions'' at distances $\leq 3.5\times10^{-13}$  cm since the rate at large angles was inconsistent with his calculation based on scattering from a point charge. People argued that this was probably because he used classical mechanics to predict his rates. Oppenheimer\cite{Oppenheimer} soon did the problem quantum mechanically and confirmed Rutherford's predictions. The discussion in this paper
is mostly focused on the possibility of a new type of short range force rather than on the size of the nucleus.
	
	 The first clear picture of the
structure of the nucleus didn't appear for another 30 years.

	Robert Hofstadter\cite{Hofstadter} developed the electron scattering method at Stanford in the mid 1950's using a $\sim 60-200 MeV$ electron accelerator.  The scattering of
relativistic electrons from an extended target differs in several respects from the form calculated by Rutherford- ie.

\begin{equation}
(\frac{d\sigma}{d\Omega})_{Rutherford}=1/4(Z\cdot\alpha _{EM})^2\frac{(\hbar c)^2}{E_e^2}Cosec(\theta/2)^4
\end{equation}

\begin{equation}
(\frac{d\sigma}{d\Omega})_{Mott}=(\frac{d\sigma}{d\Omega})_{Rutherford}\cdot Cos(\theta /2)^2(1+\frac{\pi Z \alpha _{EM} Sin(\theta /2)(1-Sin(\theta /2))}{Cos(\theta /2)^2})
\end{equation}

\begin{equation}
\rho(r)=\frac{1}{8\pi (a)^3}Exp(-r/a)
\end{equation}

\begin{equation}
FormFactor(Q)=\frac{4 \pi  \int _0^{\infty }r \rho (r,a) \sin (r Q)dr}{Q}
\end{equation}

\begin{equation}
(\frac{d\sigma}{d\Omega})_{Hofstadter}=(\frac{d\sigma}{d\Omega})_{Mott} FormFactor(Q)^2
\end{equation}
	
		where Z is the nuclear charge,$ \alpha _{EM}$ is the fine structure constant, $\hbar c$ is the reduced Planck constant times the speed of light and Q is the momentum transfer. The Form Factor suppression can be calculated from a given structure model, as in eqn. 3. Hofstadter realized that inelastic events in which the
electron transferred energy to the nucleus should be excluded in the structure measurement so he measured outgoing energy to $\sim$ O(1 MeV). He also measured the form factor of protons.

	Starting in the 60's Friedman, Kendall and Taylor extended Hofstadter's measurements of the proton using the then new Stanford 2-mile long linear accelerator. Based on an analysis of the inelasticity and scattering angle\cite{Friedman}, their measurements showed that the proton is also composite- consisting of pointlike partons (the
quarks). 

	 At roughly this point the picture of structure in nuclei and protons moved from a coordinate description to one using the momentum distribution of the constituents, as
in the experiment of Friedman et al. depicted in Figure 3. It would be interesting to construct a unique model of the proton based on the measured distributions. However, up
to now, it has only been possible to do the opposite- ie to predict the distribution starting from certain models.
\begin{figure}
\centering
\includegraphics[width=0.8\linewidth]{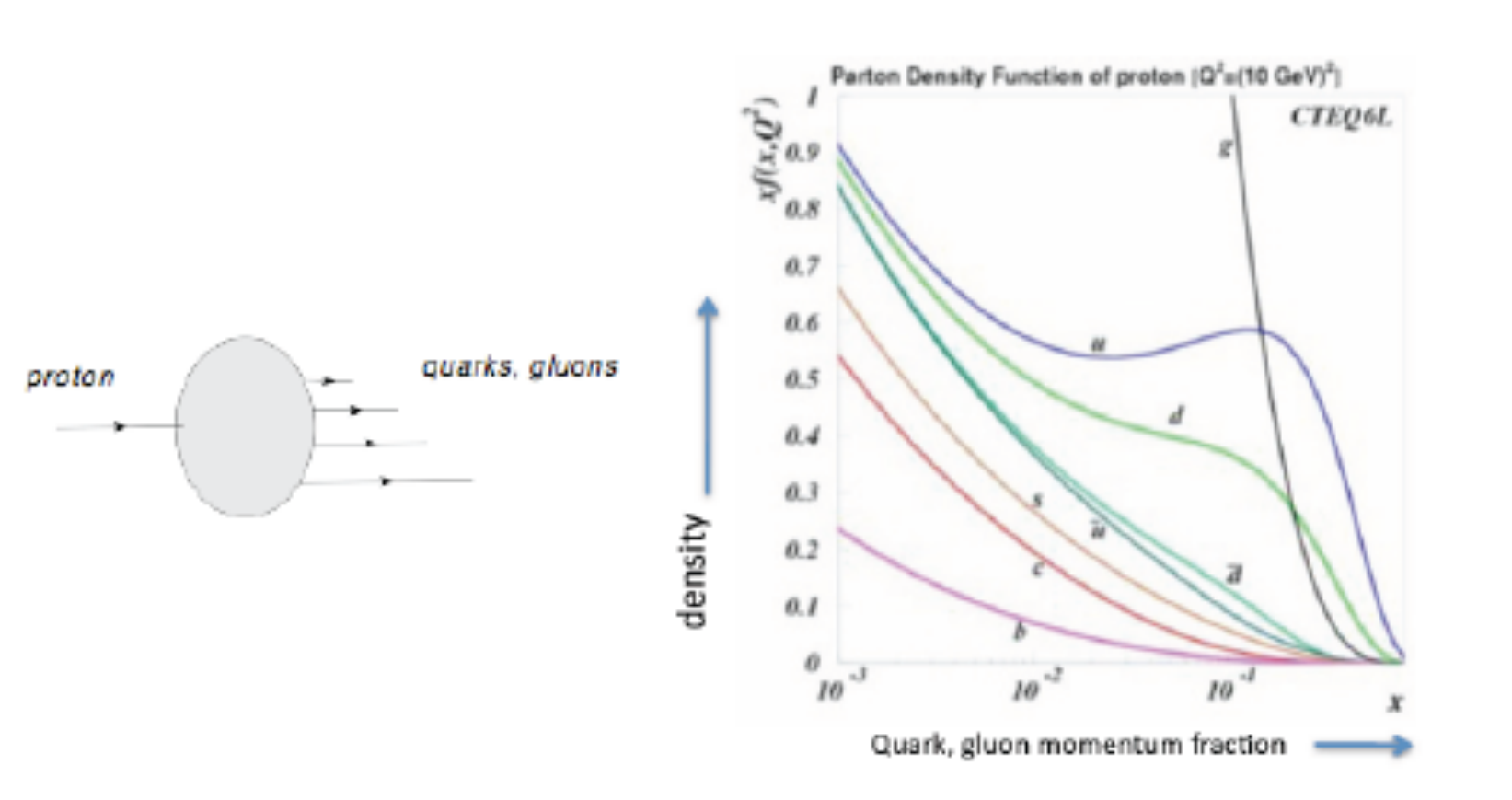}
\caption{Proton structure in deep inelastic scattering.}
\label{fig:pdf2}
\end{figure}
\begin{figure}
\centering
\includegraphics[width=0.6\linewidth]{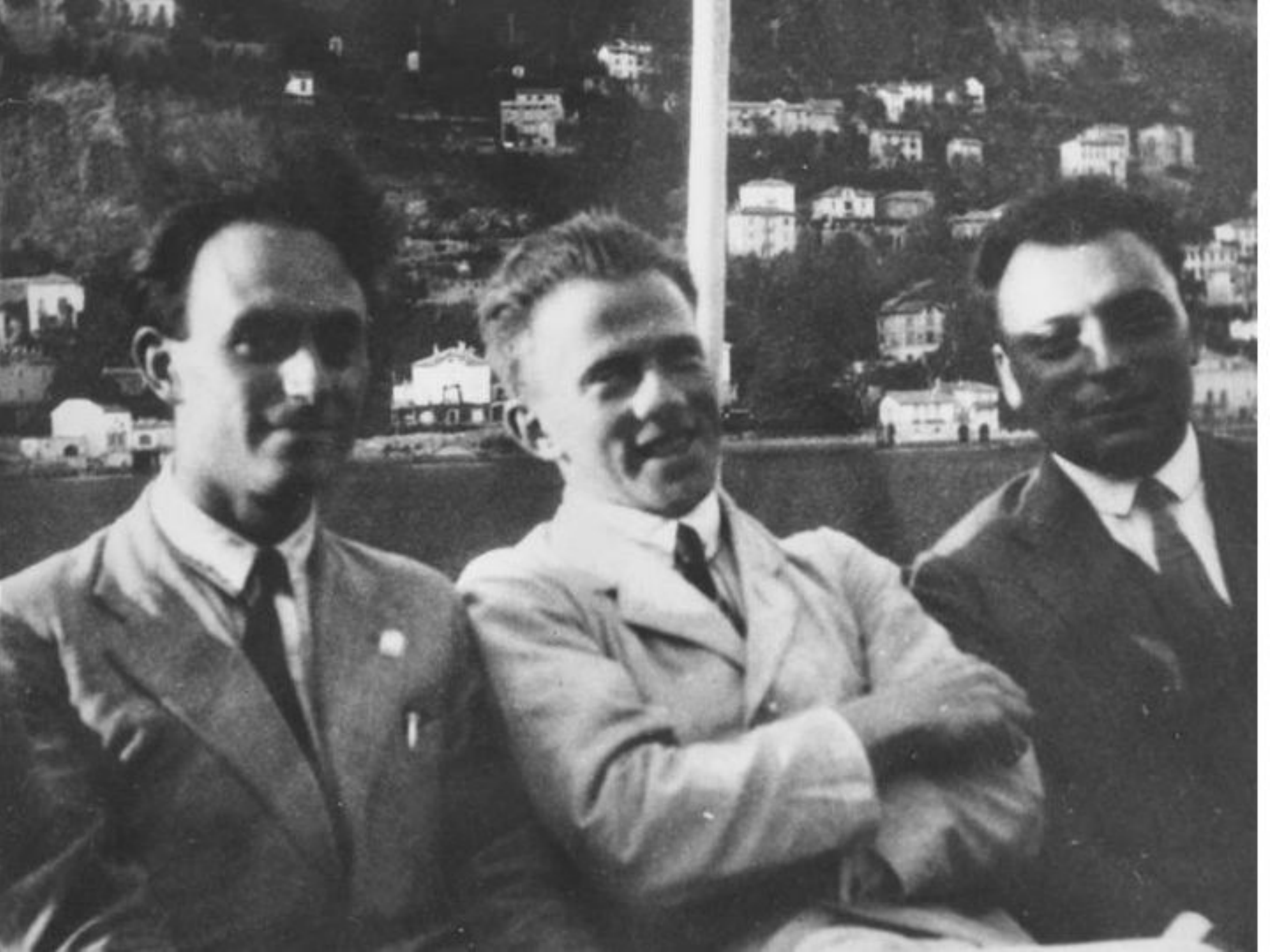}
\caption{Enrico Fermi, Werner Heisenberg and Wolfgang Pauli around the time of Fermi's paper.}
\label{fig:fermi}
\end{figure}

\section{Inelastic Scattering}
\subsection{Enrico Fermi}

	In the spring of 1924 Enrico Fermi returned to Rome after an unsatisfying period in Germany. There he met George Uhlenbeck who was tutoring the children of the Dutch 
ambassador. Uhlenbeck encouraged Fermi to go to Leyden and visit his teacher, Paul Ehrenfest. So Fermi applied to the International Education Board- a Rockefeller family charity- and received support for a 3 month stay. The Board's appraisal of Fermi and his trip report can be found in the Rockefeller archives. In his report Fermi says he 
learned many things in Leyden and wrote 2 papers including the following one\cite{Fermi}. 

\subsection{Fermi's paper}
	
	Fermi was interested in calculating the interactions of charged particles-electrons and $\alpha$-particles- with gases. Since much was known about the 
interaction of X-rays and other photons, he proposed that many of those phenomena (ie resonant absorption) should be produced by fast particles also.
\begin{figure}
\centering
\includegraphics[width=0.5\linewidth]{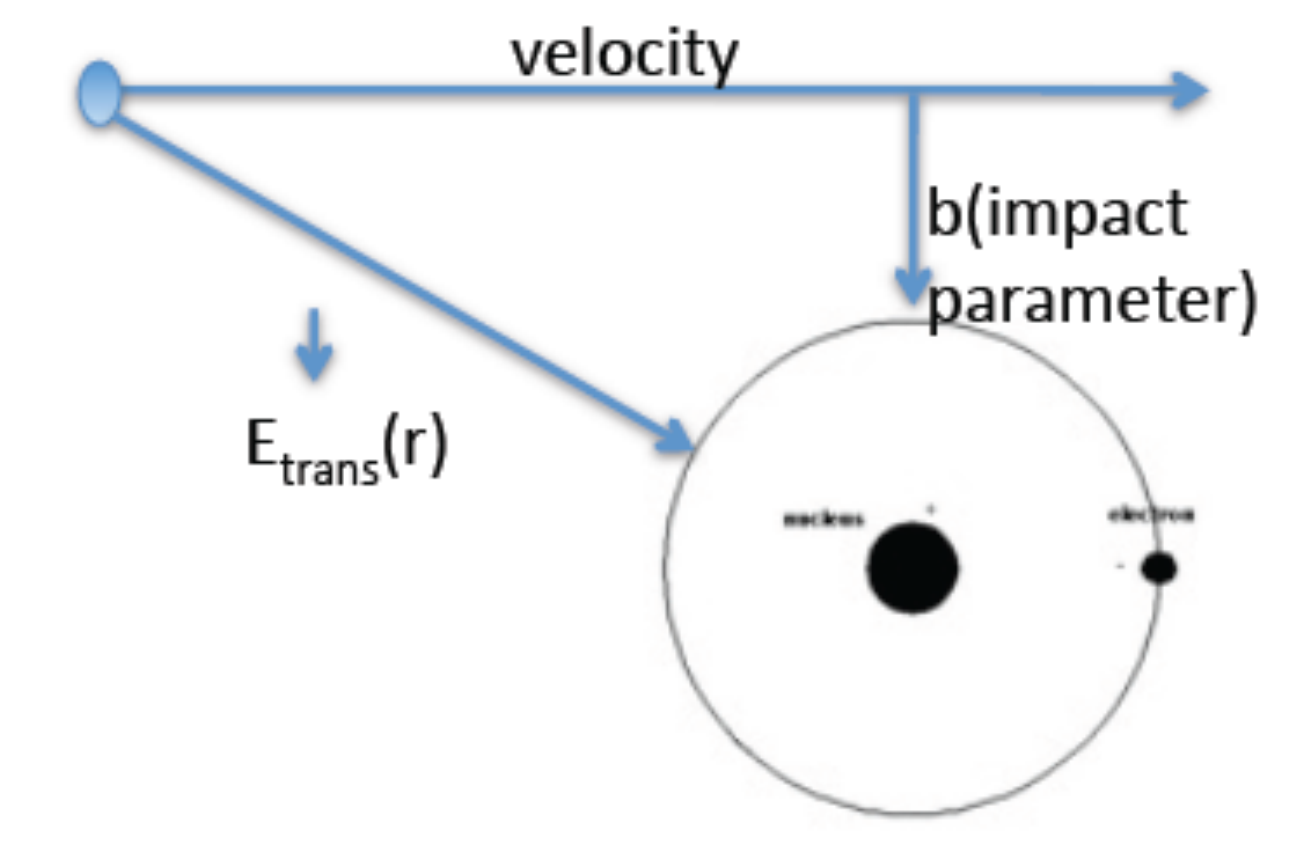}
\caption{A charged particle with velocity, v, and charge, q, moving by an atom at an impact parameter, b.}
\label{fig:epa}
\end{figure}
	So Fermi analyzed the time dependent electric field seen by an atom, which a charged particle passes with speed, v, and impact parameter, b.
The transverse component (the longitudinal component is not important for large v) can be written as

\begin{equation}
E_{trans} =\frac{ q\times b}{(b^2 + v^2 t^2 )^{3 / 2}}
\end{equation}

	which he then rewrote as an expansion in a harmonic series as
\begin{equation}
E_{trans}=\Sigma a_N^2 cos(\frac{2\pi n t}{T})
\end{equation}
	So the interactions of the charged particle are equivalent to those of a ``field of light'' with intensity $a_N^2$ at frequency $\frac{n}{T}$. For resonant excitation
all frequencies would be ineffective except at the resonant frequency.

	 In this paper Fermi also tries to check his calculation with available data so he developed a
more practical expression for his calculated interaction probability at a given impact parameter, P(b). He introduces an equivalent interaction radius of the atom
\begin{equation}
\pi\rho^2=2\pi\int_0^{\infty} b P(b) db=\sigma
\end{equation}
	ie as if P=1 for $b\leq\rho$.
	Today the term cross section ($\sigma$) is used and usually expressed in the units of barns ($10^{-24} cm^2$), which is practical since a cross section of 1 barn per atom would
give roughly 1 interaction in a typical target. For processes discussed later, cross sections range from 33,000 barn in Au+Au$\rightarrow$Au+Au+$e^-e^+$ to 0.1 barns for the
total proton-proton interaction to $10^{-14}$ barn for central exclusive Higgs production at the LHC.
	
	Fermi's paper was originally written in German and submitted to Zeitschrift fur Physik in 1924. In 1925 he also wrote an Italian version and submitted it to Il Nuovo Cimento.
Only the German version appears in his collected works so it is less read. Nevertheless Persico says that this was one of Fermi's favorite ideas and he often used it later in
his life. The German version was translated into Russian and perhaps other languages but it only recently appeared in English.

	Hunter Thompson(no relation to JJ) often stayed up nights typing out pages from ``The Great Gatsby'' just to see what it would be like if he could write like that. This paper would be worth trying.
\begin{figure}
\centering
\includegraphics[width=0.6\linewidth]{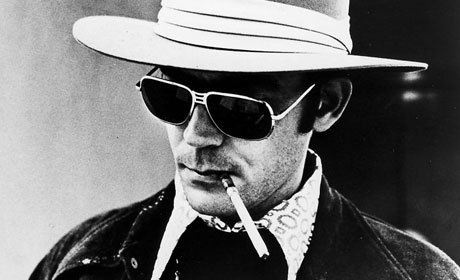}
\caption{Hunter S. Thompson.}
\label{fig:Hunter}
\end{figure}

\section{Applications of  the EPA}

	Fermi's method, the equivalent photon approximation (EPA), is often used for problems where the basic photon interaction probability is known experimentally or can be calculated. Two examples are critical for the operation of the LHC as a nuclear beam collider. 
	
	In the first one\cite{Baltz} electron pairs are created in the intense fields of the (Pb) colliding nuclei ($\sigma\propto Z^4$).
	 The equivalent photon spectrum of the nuclei is calculated and the colliding photon flux multiplies the calculated Breit-Wheeler pair photoproduction cross section-see Figure.7 . This pair production process would be harmless at a collider but in a fraction of this cross section ($\sim$250 barns at LHC) the electron is subsequently captured changing the charge of the beam nucleus by one.

\begin{figure}[h]
\centering
\includegraphics[width=0.6\linewidth]{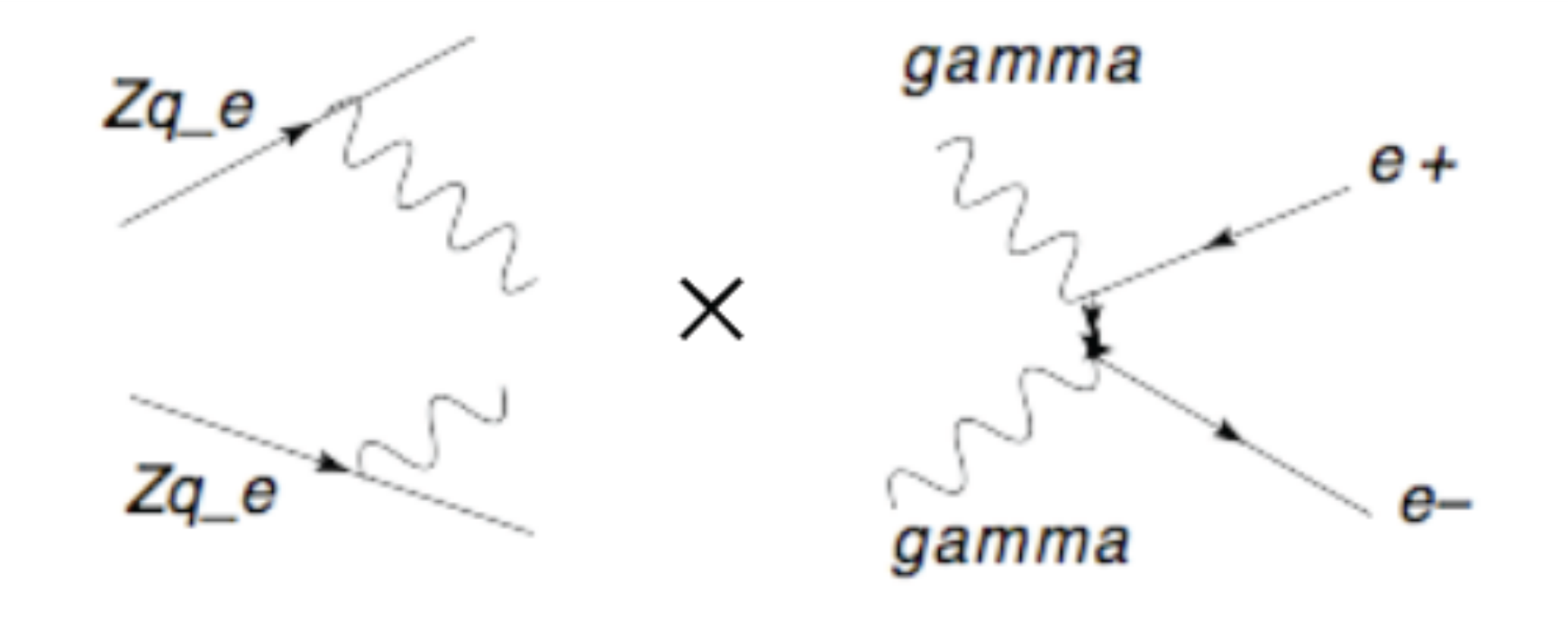}
\caption{Electron-positron pair production is calculated using the equivalent photon approximation and the Breit-Wheeler photon cross section.}
\label{fig:breit-wheeler}
\end{figure}

	The other dominant process is direct interaction of equivalent photons from one beam with nuclei in the other beam. Low energy ($\sim$ 6 MeV) photonuclear cross sections are very large and well measured. In this calculation\cite{BCW} the cross sections are weighted by the calculated photon spectrum. In these interactions a neutron is emitted which also changes the atomic mass but not the charge.
\begin{figure}[h]
\centering
\includegraphics[width=0.6\linewidth]{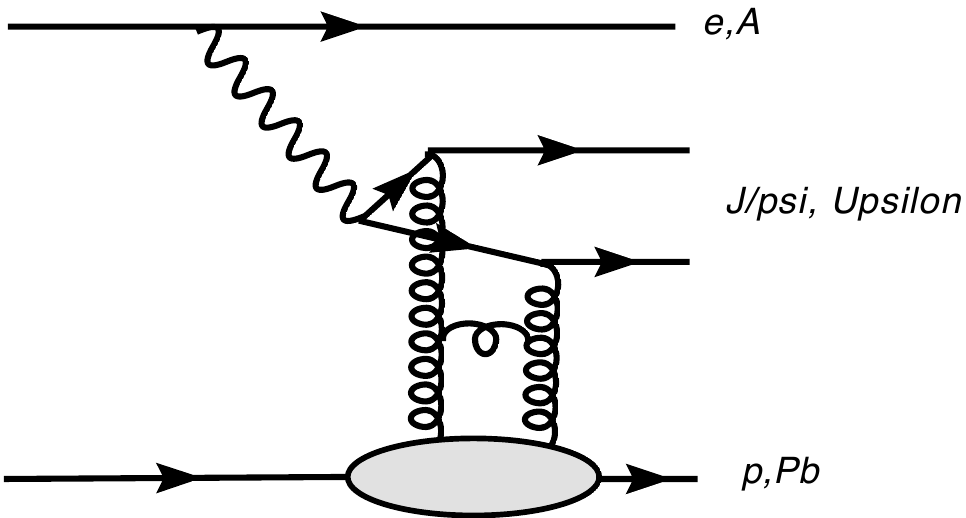}
\caption{Diffractive J/$\Psi$ photoproduction in ATLAS. The J/$\Psi$ interacts with a lead nucleus via exchange of 2 gluons.}
\label{fig:vecmeson}
\end{figure}
	
	Both processes change the trajectory of beam particles through the magnetic optics of the LHC and so some beam is lost. Beam loss is a major concern at the LHC since it leads to energy deposition into the liquid He cooled LHC magnets. These two processes limit the maximum intensity in nuclear beam operation of the LHC. The intensity limit is local (a limit per experiment) since the damage caused by beam loss is local.

	The high energy photon flux in collisions of heavy nuclei at the LHC will make possible unique measurements of nuclear and proton structure\cite{SVW,UPC}. As shown in Figure 8, a high energy photon can fluctuate into a bound state of a charmed quark-antiquark pair ( the J/$\Psi$). Since the J/$\Psi$ has the same quantum numbers as the photon,
it can scatter coherently off the entire nucleus without breaking it up. This is analogous to Hofstadter's measurement and the scattered J/$\Psi$ distribution is related
to the structure as in eqn.3. However, unlike electron scattering, this structure is measured by exchanging gluons so it images the ``gluon charge'' distribution of the 
nucleus. PHENIX made initial measurements of this process\cite{PHENIXUPC}.

	EPA has also been applied beyond electromagnetic interactions. For example, the coupling between a Higgs boson and
a pair of W bosons is known from theory. Since the coupling of the W boson to the proton is related to the $\beta$-decay rate of the neutron, the equivalent flux of W's, from colliding
protons, which fuse to form the Higgs can be calculated and used to predict Higgs production\cite{Dawson}. This is the dominant Higgs production mechanism if $M_{Higgs} \geq\sim 200$ GeV.

\section{Coherence}
	
	The photon flux from a heavy nucleus is very large over most of the spectrum because the intensity is proportional to $Z_{nucleus}^2$. However at short enough wavelengths the intensity drops rapidly since photons no longer couple to the full nuclear charge. This occurs when the wavelength is less than the nuclear size  -or $q\geq\frac{\hbar c}{2¹R_{nucleus}}$ - about 20-30 MeV.
	
	This tiny energy translates to a very large one in relativistic collisions of nuclei as in the spectrum seen
by one nucleus in its rest frame (the target) colliding with another at the LHC, where Pb beams have a relativistic factor, $\gamma$, of 3,000. In this rest frame the photon endpoint is boosted by a factor of $2\gamma^2$ so effectively the endpoint becomes 400,000 GeV. No photons of such a high energy have yet been seen in the lab
or in cosmic rays.
\begin{figure}[h]
\centering
\includegraphics[width=0.8\linewidth]{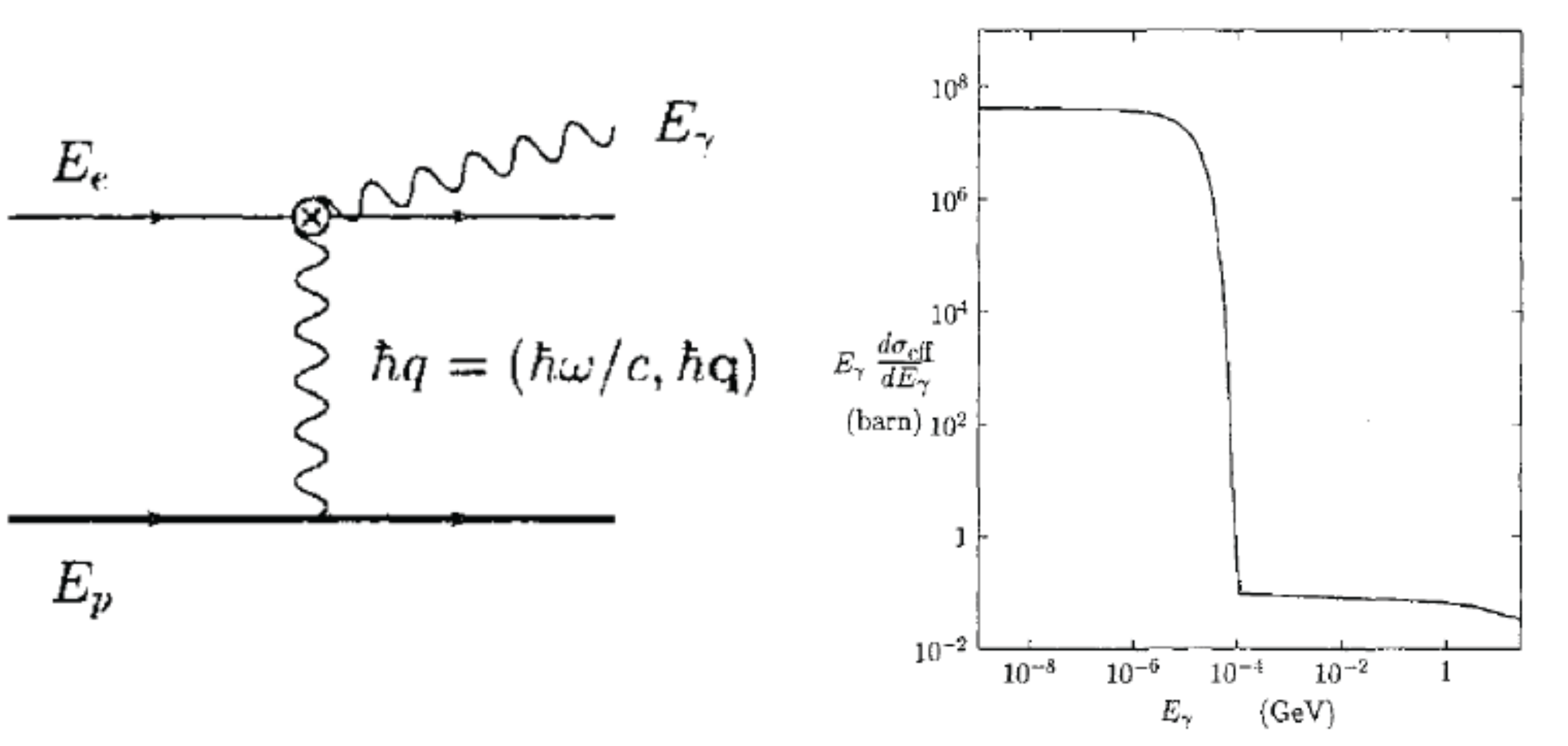}
\caption{Beamstrahlung and $K_{Short}$ regeneration are two examples in particle physics of coherence at a centimeter scale.}
\label{fig:serbo}
\end{figure}

	Perhaps the most dramatic example of the role of coherence is in a``Beamstrahlung'' calculation\cite{Serbo} for the HERA electron-proton collider in Hamburg, Germany. The idea behind this process is that when a short bunch of electrons pass through a 1 cm long bunch of $10^9$ protons, some of the electrons should scatter and radiate photons.
This radiation would be a way of imaging the proton bunch, which is useful to accelerator physicists.

	In the calculation of Serbo et al. it is found that the radiation intensity suddenly increases by 9 orders of magnitude at wavelengths corresponding to 
 $\sim$ 1 cm (the proton bunch length).
	
\section{Diffraction}

	In 1955 early measurements of deuteron collisions with heavy nuclei were used to study a process proposed by Serber- deuteron stripping.
Since the deuteron has a large radius it was expected that in some collisions only the proton in the deuteron would interact with the rim of the
target nucleus. This would be a way of producing a high energy neutron beam.

	Glauber\cite{Glauber0} showed that another type of interaction would also lead to free forward neutrons, which he called ``free dissociation''. He calculated the cross section for interactions in which both proton and neutron escaped inelastic interaction with the nucleus, but nevertheless the deuteron wavefunction
overlaps with the nucleus. This would cause transitions to other configurations of the proton-neutron system. Since the deuteron has only one bound state it will break up. He found that ``free dissociation'' is almost as large ($\sim60\%$) as the Serber cross section.

	The PHENIX experiment studied dissociation of the deuteron in relativistic d-Au collisions\cite{Glauber} and observed both the Serber process and the Glauber one through the
outgoing proton and neutron. At high energies dissociation, which has a significant contribution from photodissociation, is used as a basis for calculating other cross
sections in d-Au collisions in PHENIX.
\begin{figure}[h]
\centering
\includegraphics[width=0.3\linewidth]{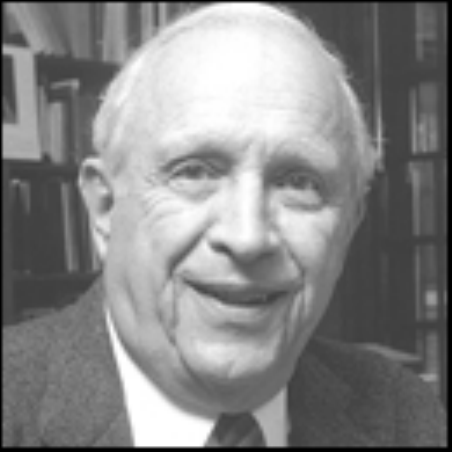}
\caption{Roy Glauber.}
\label{fig:glauber}
\end{figure}

	In the years following Glauber's paper there were others\cite{FP,GW} suggesting that the same picture could be applied to the dissociation of elementary particles like the proton at high energies. Since, in the case of the proton, the wavefunction isn't known the following picture was used for the deuteron case.
	
		In the deuteron interaction with a nuclear target we could, generally, represent the deuteron ground state in terms of configurations ranging from small ones to larger ones which each have well defined scattering amplitudes. Then the average amplitude determines elastic deuteron scattering while the dispersion in scattering amplitudes is
		proportional to dissociation.
		
		Experimentally we find very little growth in the proton diffraction dissociation cross section above ISR energies while elastic and total grow logarithmically. For this 
		reason you could say that the dispersion is shrinking with energy.
		
	A surprising aspect of  dissociation data on protons, mesons and photons\cite{E612} and nuclei\cite{Willis}  , accumulated over the next years, was how large the dissociation cross section is-
particularly at higher energies. The same coherence condition discussed above for electromagnetic interactions was found to hold. Since the proton size is $\sim\frac{1}{m_{\pi}}$, we expect to see the cross sections increase rapidly above this wavelength-- which they do. You could ask ``the wavelength of what?''. The process, which came to be called ``diffraction dissociation'' had too large a cross section to be electromagnetic and didn't correspond to the expected behavior due to exchanges of known particles.

\section{Hard Diffraction}

	In 1985 Ingelman and Schlein\cite{IS} proposed that one way to get at the properties of the exchanged system in diffraction (the Pomeron) was to probe it using hard collisions of 
e-proton and proton-proton. The Pomeron would have to be a component of the proton constituent flux. Hard e-proton collisions which leave the proton in tact then result
from electron scattering off the Pomeron's constituents. So these measurement would answer questions about the Pomeron structure- for example, whether it is made up of quarks.

	H1 and ZEUS(e-proton) and CDF($p\bar{p}$) studied hard scattering, such as W production, in non-diffractive and diffractive (where the proton remains in tact) scattering. Aspects of their results disfavored this Pomeron picture. For example, the ratio of diffractive to non-diffractive production in a number of hard processes in CDF\cite{Rock} was found to be of order 1$\%$ while it was $\sim 10\%$ in e-proton scattering. If the Pomeron flux is an attribute of the proton this fraction should have been the same in both cases.
	
	CDF also found a significant cross section for hard production, such as high transverse energy jets, in diffractive interactions where both $\bar{p}$, (detected downstream), and the proton (inferred from the absence of other particles at forward angles) remain intact. This process is analogous to Breit-Wheeler electron pair production in Figure 7 with photons replaced by the Pomeron. In a large sample of these events CDF found that the hard scatter is exclusive.

\begin{figure}[h]
\centering
\includegraphics[width=0.5\linewidth]{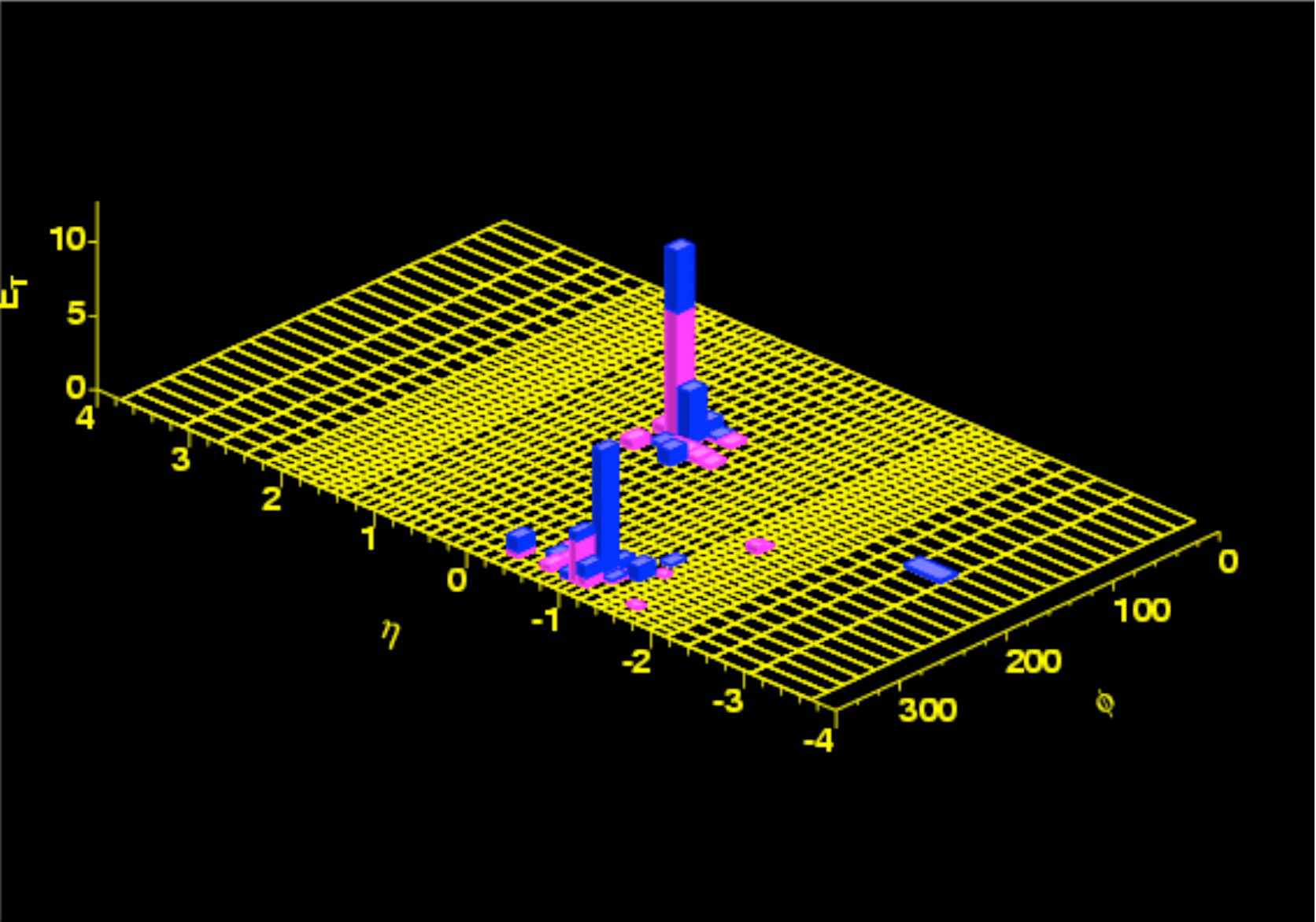}
\caption{One of several thousand CDF central exclusive production events.  The distribution of energy in $\eta= -ln$ $tan(\frac{\theta}{2})$ is very concentrated near $90^o$ and no proton remnants are seen.}
\label{fig:cdf}
\end{figure}
	
\section{Central Exclusive Production}

	If the Higgs Boson can be observed in the exclusive process found by CDF it would be very useful, even if the Higgs is found earlier in inclusive searches. In the CDF events the forward p ($\bar{p}$) can be used to calculate the energy that should be seen in the central system (ie the 2 jets in figure 11) based on the momentum loss of the p ($\bar{p}$) . Using this missing mass technique in the ATLAS experiment, the uncertainty in the measurement of the Higgs mass would be $\sim 2\%$ independent of the decay modes of the Higgs. 
	
	Helicity conservation in exclusive production is also very useful. If the Higgs is produced in this way then it very likely has the expected quantum numbers of $0^{++}$. Because Higgs decay to 2 heavy quarks has a significant background from direct production not involving Higgs decay it's important that helicity conservation suppresses this background by a factor of $(m_{quark}/m_{Higgs})^2$ ie$\sim 2\times10^{-3}$. CDF sees roughly the expected b-quark suppression in exclusive production. For all of these reasons diffractive Higgs production is an important complement to inclusive searches.

	It is difficult to calculate diffractive Higgs production\cite{Khoze}, partly because such a hard interaction requires collisions at small impact parameters where absorption is significant.
The following figure from Khoze et al. shows the elementary process in exclusive production (ie not including absorptive corrections). Note that a second, soft gluon exchange is needed to screen color charge and
to ensure the protons remain in tact. This is very different from the Pomeron exchange picture or EPA since it doesn't factorize into a flux term and an interaction term.
\begin{figure}[h]
\centering
\includegraphics[width=0.5\linewidth]{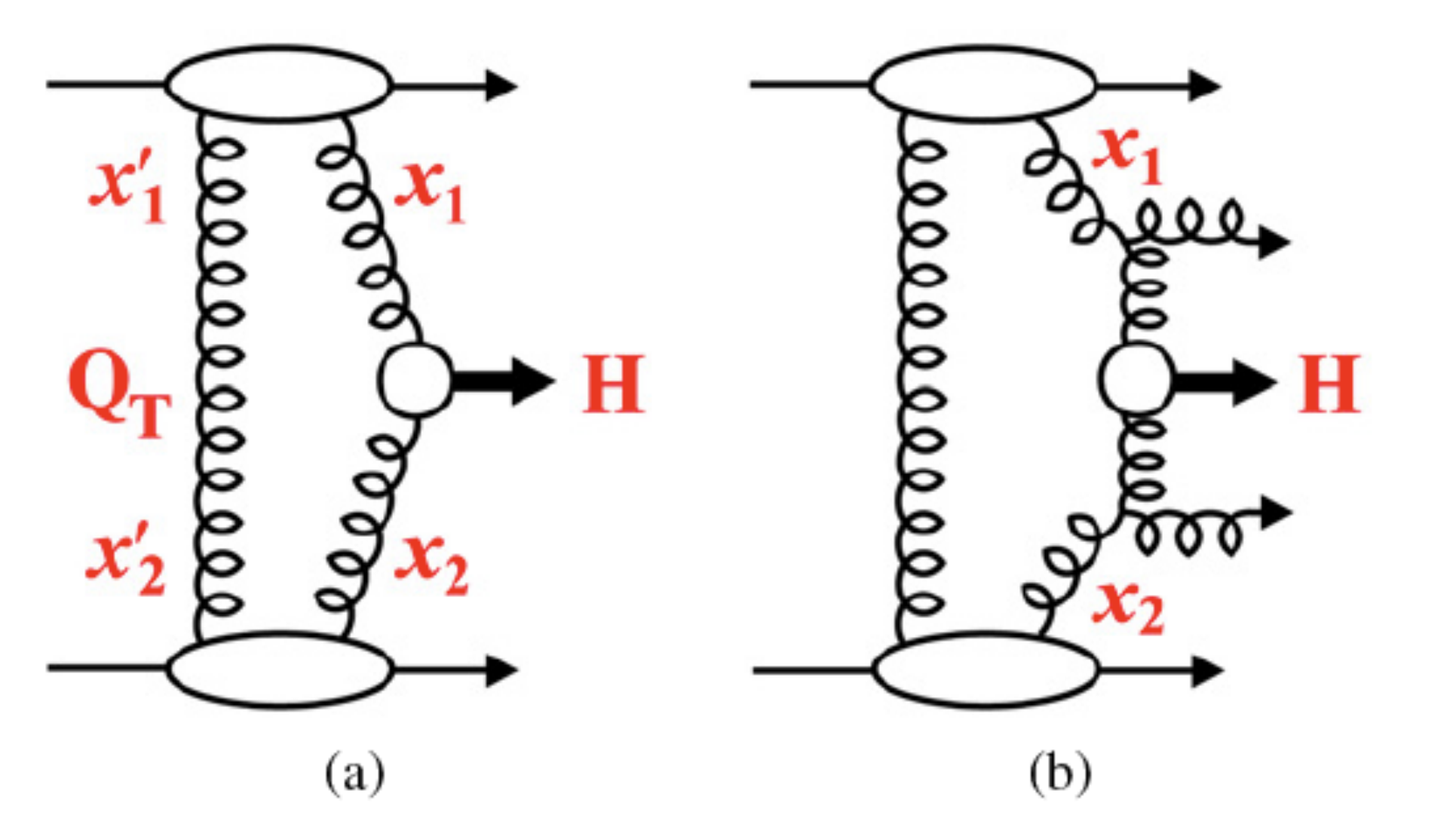}
\caption{Central Exclusive Higgs production(a) and a competing diagam (b) involving additional radiation.}
\label{fig:cdf}
\end{figure}
	The calculation of Khoze, Martin and Rhyskin was used to predict the CDF exclusive di-jet rate and agreed well. Nevertheless, there are uncertainties in 
extrapolation to LHC energies related to absorption effects. In their calculation this effect- the gap survival probability- is $\sim 2-3\%$, while some recent calculations find a smaller value, $\sim 1 \%$. So the current uncertainty in exclusive Higgs production is about a factor of 2. In the standard model the 
prediction is $\sim 0.5-1\times10^{-14}$ barn, while in the MSSM model the cross section could be 10 times higher.

\section{Instrumentation}

	At design performance the LHC will store beams of a few$\times10^{14}$ protons with energy of 7,000 GeV. This adds up to about 400 MegaJoules or, roughly,
the kinetic energy equivalent of a couple locomotives at top speed. A great deal of care is required to ensure that even a small fractional loss of the beam is prevented.
One measure is to insert collimators close to the beam. The beams can be as small across as 0.02 mm so there is getting to be useful expertise locating things near the beam.

	ATLAS is a large detector. It would fill half  the Notre Dame cathedral in Paris and is instrumented with the equivalent of 100 Megapixels of sensors. Because of the bunch structure of the beam at the LHC, the 1 Gigahertz interaction rate is clustered in 40 Megahertz bursts with a time spread of 170 picoseconds. So the ATLAS detector
has effectively a 40 Megahertz frame rate with a typical pileup of 25 events per frame. Some of the detector systems also record the event time to better than 170 picoseconds
so time could be used to associate their information with a particular event.

	Downstream of the main detector proton tracks from diffractive events, useful for the Higgs measurement, would separate from the core of the beam by about 0.3 to 1.5 cm at locations 220m and 420m away. Measuring 1 proton 3 mm away from $10^{11}$ protons is challenging but seems to be doable. What is more difficult is to associate protons with the correct event of 25 recorded in the same frame\cite{SNW,FP420}. The only possible tool for this is timing and, since the mean separation between events is $\frac{2\times 170}{25}$= 13.6 picoseconds, the necessary timing precision is of this order. Light goes 3 mm in 10 picoseconds. There is currently no particle detector that can measure
10 picoseconds at rates of 10 MHz.

	We have started a generic program on related devices that could achieve this performance\cite{APD}. If a charged particle goes through a deep depleted avalanche photodiode(APD) it will produce up to several thousand e-hole pairs in the silicon, which are then amplified internally by factors up to several hundred. This signal is much faster than the ones in
photodetection (for which APDs are normally used). It seems likely, based on our tests, that one of our devices will be suitable for ATLAS.

	A pre-production device from Hamamatsu that we are evaluating is also based on APDs and should fit our requirements. In this device- a Hybrid Avalanche Photodiode(HAPD)-
photons (from a Cerenkov radiator) hit a transparent photocathode and produce photoelectrons, which are accelerated to 8 kiloVolts and electrostatically focused on a small area APD target. This target can be much faster than the other APDs we are testing. For example, INTEL has announced a new APD with 340 GigaHertz gain-bandwidth product\cite{INTEL}.
Those devices are very small but this is less of a limitation in a hybrid APD. The HAPD would have a major advantage in rate capability. For example, tests show that it will have a lifetime of $\geq 250$ Coulomb/$cm^2$ which is more than 1,000 times better than Microchannel plate photomultipliers. Using a femtosecond laser we obtained a
single photon response of $\sim 11$ picosecond\cite{Nano}.

	It's sometimes said that, compared to the scientific benefits, the applications of LHC technology could have a more immediate impact. This could certainly be true of
10 picosecond timing. One example would be medical imaging, where positron emission tomography(PET) has still limited clinical application, even though it is more elegant than
SPECT. Since the range of $\beta$-decay positrons in tissue is typically less than 2 mm, 10 picosecond timing, if it's possible in PET, would be transformative since the position would be measured in each decay event at the ultimate resolution. This would imply reduced doses or even real time imaging of events at the blood-brain interface.

\section{Acknowledgements}

	I'm grateful to Dino Goulianos, Mark Strikman and Genya Levin for helpful discussions about diffraction and to Brian Cox and Robin Marshall for interesting discussions about Rutherford. I thank the organizers- particularly Alejandro Ayala- for a very stimulating meeting.


\begin{thebibliography}{10}
\bibitem{Rutherford}
E.~Rutherford and J.~Chadwick, ``The Scattering of $\alpha$-particles by Helium'', Philosophical Magazine Series 7, 4:22, 605-620.
\bibitem{Oppenheimer}
J.~Oppenheimer, Zeitschrift fur Physik, xliii p.413 (1927).
\bibitem{Hofstadter}
R. ~Hofstadter, ``The Electron-Scattering Method and Its Application to the Structure of Nuclei and Nucleons'', Nobel Lecture December 1961.
\bibitem{Friedman}
J.~Poucher et al, ``High-Energy Single-Arm Inelastic e-p and e-d Scattering at 6 and $10^o$'',
Phys. Rev. Lett. 32, 118Ð121.
\bibitem{Fermi}
E.~Fermi, ``On the Theory of Collisions between Atoms and Electrically Charged Particles''. arxiv.org/abs/hep-th/0205086.
\bibitem{Baltz}
A. J. Baltz, M. J. Rhoades-Brown, J. Weneser, Phys.Rev.A50:4842-4853,1994
A. J. Baltz, Phys.Rev.Lett.78:1231-1234,1997.
Anthony J. Baltz, Phys.Rev.C71:024901,2005 and references therein.
\bibitem{BCW}
A.~Baltz,C.~Chasman,S.~White,
Nucl.Instrum.Meth.A417:1-8,1998
\bibitem{UPC}
A.~Baltz et al, ``The Physics of Ultraperipheral Collisions at the LHC'',
Physics Reports 458, nos1-3 (2008)
\bibitem{SVW}
M. ~Strikman, R. ~Vogt, and S. ~White, Phys. Rev. Lett. 96, 082001 (2006).
arXiv:hep- ph/0508296
 \bibitem{PHENIXUPC}
  PHENIX Collaboration,  Phys. Lett. B 679, 321 (2009).
  \bibitem{Dawson}
  S. ~Dawson, ``The Effective W Approximation'', Nucl.Phys. B249 (1985) 42.
  \bibitem{Serbo}
  V.M. ~Budnev, I.F. ~Ginzburg, G.V. ~Meledin, V.G. ~Serbo, Phys. Rept. 15(1974)181
  \bibitem{Glauber0}
  R.J.~Glauber, ``Deuteron Stripping Processes at High Energies'', Physical Review 99 no. 5 p.1515 (1955).
  \bibitem{Glauber}
  S.~White, ``Diffraction Dissociation- 50 Years Later'', DIS05 Proceedings.
  arXiv:nucl-ex/0507023v2
  \bibitem{FP}
  E.L.~Feinberg and I.~ Pomeran$\check{c}$uk, ``High Energy Inelastic Diffraction Phenomena'', Supplement to Volume II, Series X
  of Il Nuovo Cimento No. 4 (1956). This paper presents a survey of theoretical work with a main focus on electromagnetic phenomena
  such as coherent bremstrahlung. It also presents a calculation of deuteron dissociation. I thank Mark Strikman for pointing out an 
  earlier paper- not translated from the Russian: I.Ja.~ Pomeran$\check{c}$uk and E.L.~Feinberg, ``Diffractive Regeneration of Particles
  in Nuclear Collisions'', Dokl. Akad. Nauk SSSR, 93 439 (1953).
  \bibitem{GW}
  M.L.~Good and W.D.~Walker, ``Coulomb Dissociation of Beam Particles'', Physical review 120 No.5 p.1855 (1960) and
    M.L.~Good and W.D.~Walker, ``Diffraction Dissociation of Beam Particles'', Physical Review 120 No.5 p.1857 (1960).
 \bibitem{E612}
 T.J.~Chapin et al, ``Diffraction Dissociation of Photons on Hydrogen'', Phys.Rev.D31:17-30,(1985)
 \bibitem{Willis}
 W.J.~Willis et al,``Diffraction dissociation of nuclei in 450 GeV/c proton-nucleus collisions'', Zeitschrift fur Physik, 49 no.3 (1990).
 \bibitem{IS}
 G. ~Ingleman and P. ~Schlein, Phys. Lett. B 152, 256 (1985).
 \bibitem{Rock} 
 All of the CDF diffractive papers can be found at: http://physics.rockefeller.edu/publications.html.
 \bibitem{Khoze} see eg. V.A. ~Khoze, A.B. ~Kaidalov, A.D. ~Martin, M.G. ~Ryskin, W.J. ~Stirling, ``Diffractive processes as a tool for searching
for new physics"; hep-ph/0507040
 \bibitem{SNW}
 S.N.~White, ``On the Correlation of Subevents in the ATLAS and CMS/Totem Experiments'', arxiv.org/pdf/0707.1500
 \bibitem{FP420}
 ``The FP420 R$\&$D Project'': www.fp420.com/papers/fp420.pdf
 \bibitem{APD}
 S.N.~White et al, ``Design of a 10 picosecond Time of Flight Detector using Avalanche Photodiodes'', arxiv.org/pdf/0901.2530
 \bibitem{INTEL}
 ``The Next Silicon Photonics Breakthrough'', techresearch.intel.com/articles/Tera-Scale/1612.htm
 \bibitem{Nano}
 T.~Tsang and S.N.~White, ``High-speed hybrid photodetector in single-photon counting'', presented at 5th Workshop on Making Single Molecule
 Fluorescence Lifetime Measurement Simple.
\end{thebibliography}
\end{document}